\documentclass{article}

\usepackage{arxiv}

\usepackage[utf8]{inputenc} % allow utf-8 input
\usepackage[T1]{fontenc}    % use 8-bit T1 fonts
\usepackage{hyperref}       % hyperlinks
\usepackage{url}            % simple URL typesetting
\usepackage{booktabs}       % professional-quality tables
\usepackage{amsfonts}       % blackboard math symbols
\usepackage{nicefrac}       % compact symbols for 1/2, etc.
\usepackage{microtype}      % microtypography
\usepackage{lipsum}		% Can be removed after putting your text content
\usepackage{graphicx} 
\usepackage{amsmath}
\usepackage{gensymb}
\usepackage{amssymb}
\usepackage{setspace}  % Needed for Double Spacing
\usepackage[english]{babel}
\usepackage[]{natbib}
\usepackage{siunitx}

\newcommand{\aap}{    {\it Astron. Astrophys.}}

\newcommand{\apj}{    {\it Astrophys. J.}}
\newcommand{\apjl}{   {\it Astrophys. J. Lett.}}

\newcommand{\jgr}{    {\it J. Geophys. Res.}}

\newcommand{\solphys} {{\it Solar Phys.}}

\newcommand{\ssr}     {{\it Space Sci. Rev.}}

\newcommand{\apjs}		{{\it Astrophys. J. Suppl.}}

\title{Forecasting the Ambient Solar Wind with Numerical Models. II. An Adaptive Prediction System for Specifying Solar Wind Speed Near the Sun}

\date{} 					% Or removing it

\author{
  Martin A.~Reiss \\
  NASA Goddard\\
  Greenbelt, MD 20771, USA \\
  \texttt{martin.reiss@oeaw.ac.at} \\
  \AND
  Peter J.~MacNeice \\
  NASA Goddard \\
  Greenbelt, MD 20771, USA \\
  \And
  Karin Muglach \\
  NASA Goddard \\
  Greenbelt, MD 20771, USA \\
  \And
  Charles N. Arge \\
  NASA Goddard\\
  Greenbelt, MD 20771, USA \\
  \And
  Christian M\"ostl \\
  Space Research Institute \\
  8042 Graz, Austria \\
  \And
  Pete Riley \\
Predictive Science Inc. \\ San Diego, CA 92121, USA
  \And
  J\"urgen Hinterreiter\\
  Space Research Institute \\
  8042 Graz, Austria \\
    \And
  Rachel Bailey\\
  Austrian Academy of Sciences \\
  8042 Graz, Austria \\
    \And
  Mathew J.~Owens\\
Department of Meteorology \\
  University of Reading \\
  Reading, UK \\
  \And
  Tanja Amerstorfer\\
  Space Research Institute \\
  8042 Graz, Austria \\
    \And
  Ute Amerstorfer\\
  Space Research Institute \\
  8042 Graz, Austria \\
}

\begin{document}
\maketitle

\vspace{-0.7cm}
\begin{abstract}
The ambient solar wind flows and fields influence the complex propagation dynamics of coronal mass ejections in the interplanetary medium and play an essential role in shaping Earth's space weather environment. A critical scientific goal in the space weather research and prediction community is to develop, implement and optimize numerical models for specifying the large-scale properties of solar wind conditions at the inner boundary of the heliospheric model domain. Here we present an adaptive prediction system that fuses information from in situ measurements of the solar wind into numerical models to better match the global solar wind model solutions near the Sun with prevailing physical conditions in the vicinity of Earth. In this way, we attempt to advance the predictive capabilities of well-established solar wind models for specifying solar wind speed, including the Wang-Sheeley-Arge (WSA) model. In particular, we use the Heliospheric Upwind eXtrapolation (HUX) model for mapping the solar wind solutions from the near-Sun environment to the vicinity of Earth. In addition, we present the newly developed Tunable HUX (THUX) model which solves the viscous form of the underlying Burgers equation. We perform a statistical analysis of the resulting solar wind predictions for the time 2006--2015. The proposed prediction scheme improves all the investigated coronal/heliospheric model combinations and produces better estimates of the solar wind state at Earth than our reference baseline model. We discuss why this is the case, and conclude that our findings have important implications for future practice in applied space weather research and prediction.
\end{abstract}

% keywords can be removed

\section{Introduction} \label{sec:introduction}
The evolving ambient solar wind flow and the magnetic field embedded within it are driven by the Sun's magnetic field. Thus, studying the magnetic field configuration in the solar atmosphere is of crucial importance for improving the understanding and ultimately the prediction of space weather from Sun to Earth. The magnetic configuration of open magnetic field lines, along which solar wind flows are accelerated to supersonic speeds, is important for predicting key properties in the interplanetary medium alongside the solar wind bulk speed, magnetic field strength, and field orientation. A considerable amount of literature has been published on reconstructing the global coronal magnetic field from photospheric magnetic field measurements due to the difficulty of mapping the magnetic field in the solar corona. The most popular techniques of this type are potential field source surface~\citep[PFSS;][]{altschuler69, schatten69}, nonlinear force-free field~\citep[NLFF; see, for example,][]{schrijver06}, and magnetohydrodynamic~\citep[MHD;][]{riley11c} models, as recently reviewed by~\citet{mackay12}. 

To seamlessly simulate the dynamics of the evolving ambient solar wind from the Sun to Earth, it is important to treat the photosphere, corona, and inner heliosphere as a coupled system. Therefore, operational systems for predicting the state in the evolving ambient solar wind rely on the coupling of magnetic models of the corona with models of the inner heliosphere~\citep{riley01, lee08}. The coupled coronal-heliospheric modelling system spans the range from 1~solar radii ($R_0$) to 1~au, where the coronal modeling domain typically spans the range from 1~$R_0$ to $2.5 \,R_{0}$ (PFSS) or $30 \,R_{0}$ (MHD), and the heliospheric domain spans the range from $5$--$30 \,R_{0}$ to 1~au using the global solar wind model solution from the magnetic model of the corona as an inner boundary condition. Currently, the most commonly used three-dimensional numerical MHD codes to derive stationary solutions for the ambient solar wind in the heliosphere are the Magnetohydrodynamics Algorithm outside a Sphere~\citep[MAS;][]{linker99, mikic99}, Enlil~\citep{odstrcil03}, the Space Weather Modeling Framework~\citep[SWMF;][]{toth05}, and the recently developed European Heliospheric Forecasting Information Asset~\citep[EUHFORIA;][]{pomoell18}. The three-dimensional MHD model solutions are characterized by closed magnetic field lines confining the solar wind plasma and open field lines along which solar wind flows accelerate to supersonic speed and propagate into the heliosphere.

Since the major discovery by~\cite{wang90} of an empirical relationship between the configuration of open magnetic field lines and the solar wind state measured in the vicinity of Earth, a key scientific goal in the space weather research and prediction community is to develop and optimize empirical techniques for specifying the large-scale properties of the solar wind solution at the inner boundary of the heliospheric model domain~\citep[e.g.,][]{arge00, riley01,arge04}. As the spatial and temporal structure in the wind flow is determined by the dynamic pressure term in the momentum equation, the solar wind bulk speed at the inner boundary of the heliospheric domain dominates the propagation dynamics of the evolving ambient solar wind flow in the interplanetary medium~\citep{riley15}. Empirical techniques for specifying solar wind speed near the Sun are thus critical in solar wind models such as the Wang-Sheeley~\citep[WS;][]{wang90}, Distance from the Coronal Hole Boundary~\citep[DCHB;][]{riley01}, and Wang-Sheeley-Arge model~\citep[WSA;][]{arge03}. These empirical formulae either rely on the amount by which a magnetic flux tube expands between the solar surface and a given reference height in the corona (WS), the minimum angular distance of an open magnetic field footpoint from a coronal hole boundary (DCHB), or a combination of the two (WSA).

Although the relative importance of the areal expansion factor and the minimum angular distance for specifying the characteristics of the magnetic field configuration in the solar corona is still under debate~\citep[see,][]{riley15}, the WSA model has become one of the workhorse models in the space weather community~\citep[e.g.,][]{sheeley17}. The coupled WSA/Enlil model is now routinely used by the Space Weather Prediction Center at the National Oceanic and Atmospheric Agency, and the Met Office Space Weather Operations Centre for operational space weather predictions of the solar wind state in the interplanetary medium and the arrival of coronal mass ejections at Earth. Over the last decade, the WSA relation for specifying solar wind conditions near the Sun has been one of the most frequently used modelling approaches for studying the consequences of evolving space weather in the heliosphere. Examples include the prediction of high-speed solar wind streams~\citep{owens05, macneice09a,reiss16}, the prediction of arrival time and speed of coronal mass ejections~\citep{taktakishvili09, wold18, riley18, verbeke19}, the study of the sensitivity of CME events to model parameter settings~\citep{taktakishvili10, cash15}, the propagation of coronal mass ejections in the evolving ambient solar wind~\citep{mays15,scolini19}, the prediction of solar energetic particles~\citep{macneice11, luhmann17,wijsen19}, the understanding of how the evolving ambient solar wind flow interacts with planetary magnetospheres~\citep{dewey15}, or the study of Forbush decreases in the flux of galactic cosmic rays~\citep{winslow18}. 

A considerable amount of literature has been published on validating ambient solar wind models such as the coupled WSA/Enlil model with in situ measurements at Earth~\citep[see, for instance,][]{owens08, jian11, owens13b, devos14}. These studies show that the predictive abilities of operational solar wind models are, if at all, only slightly better than a baseline model of recurrence assuming that conditions in the solar wind will persist after each synodic rotation of the Sun~\citep[e.g.,][]{riley15}. Advances in predicting the solar wind state in interplanetary space thus are of key importance for driving innovation in applied space weather research and prediction~\citep[see,][]{schrijver15,opgenoorth19}. Here we present an adaptive model system that aims to improve the predictive capabilities of models of the evolving ambient solar wind by optimizing the model settings. To this end, we propose an adaptive predictive system that fuses information from in situ measurements of the solar wind from the previous Carrington Rotation (CR) into numerical models to align the global solar wind model solutions near the Sun with prevailing physical conditions in the near-Earth space.

\begin{figure}
\begin{center}
\includegraphics[width=0.99\columnwidth]{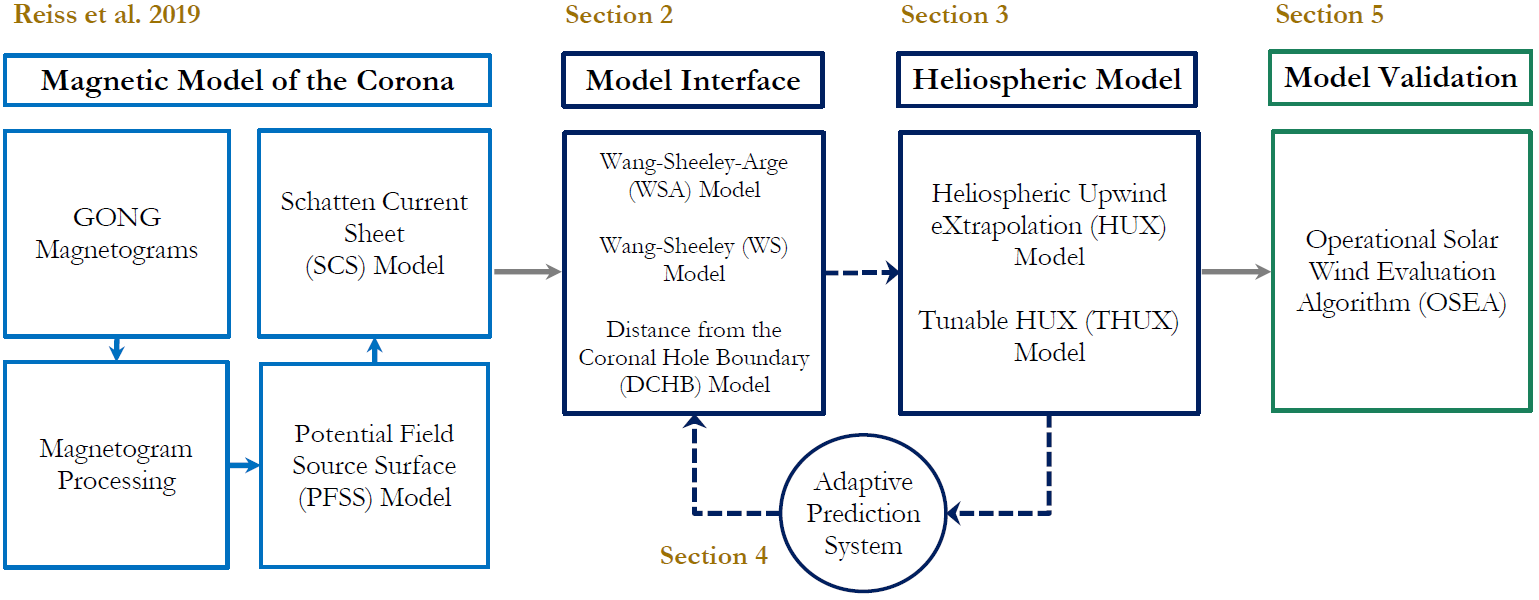}
\end{center}
\caption{Overview of the approaches used in this study illustrating the adaptive system for specifying the solar wind conditions near the Sun. The sections explaining the corresponding components are indicated.
\label{fig:f1}}
\end{figure}

Figure~\ref{fig:f1} lays out the basic steps of the proposed strategy in detail. We use the numerical framework for modelling the ambient solar wind as recently discussed in~\citet{reiss19}. Specifically, we study different empirical techniques (WS, DCHB, and WSA model) for specifying the large-scale properties of solar wind conditions at the inner boundary of the heliospheric model. To deduce the optimum model coefficient settings, we propose an adaptive prediction system to couple models of the ambient solar wind with computationally efficient heliospheric propagation tools such as the Heliospheric Upwind eXtrapolation~\citep[HUX;][]{riley11b} model and the newly developed Tunable HUX model. By doing so, we study an ensemble of possible solar wind solutions, quantitatively estimate the model uncertainties and deduce confidence boundaries. This paper is divided into the following main sections. In Section~\ref{sec:s2}, we outline the components of the numerical framework as described in~\cite{reiss19}. In Section~\ref{sec:s3}, we present the proposed modifications to the HUX model for propagating solar wind streams from the Sun to the vicinity of Earth. In Section~\ref{sec:s4}, we present the adaptive prediction system for specifying solar wind conditions near the Sun. In Section~\ref{sec:s5}, we show a detailed validation analysis of our solar wind solutions for the period 2006--2015 using the Operational Solar Wind Evaluation Algorithm~\citep[OSEA;][]{reiss16}; and in Section~\ref{sec:s6}, we conclude with a summary of the results, discuss their implications, and outline how the proposed strategies can be used effectively in operational prediction systems. 

\section{Modeling Approach} \label{sec:s2}

This section is concerned with summarizing the main components of the numerical framework for operating, validating, and optimizing models of the evolving ambient solar wind as discussed in greater depth in \cite{reiss19} and references therein. We use magnetic maps of the photospheric field from the Global Oscillation Network Group (GONG) from the National Solar Observatory (NSO) as an inner boundary condition to derive the global coronal magnetic field (see, Figure~\ref{fig:f1}).  The magnetic maps measured in Gauss (G) are given on the $\sin (\theta)$-$\phi$ grid with $180 \times 360$ grid points, where $\theta \in [0,\pi]$ and $\phi \in [0,2\pi]$ are the latitude and longitude coordinates, respectively. They are available as near real-time magnetic maps or full CR maps at the GONG online platform.\footnote{\url{https://gong2.nso.edu/archive/patch.pl?menutype=s}} 

The magnetic model of the corona couples the PFSS model and the Schatten current sheet model~\citep[SCS;][]{schatten71} to reconstruct the global topology of the solar magnetic field. The PFSS model~\citep{altschuler69, schatten69} computes the potential coronal field with an outer boundary condition such that the magnetic field is radial ($B_\theta = B_\phi = 0$) at a reference sphere, commonly known as the source surface. By forcing the magnetic field to become radial at the source surface, we simulate the effect of the solar wind flow in dragging out magnetic field lines. Beyond the PFSS model, we use the SCS model to account for the latitudinal invariance of the radial magnetic field component as observed by Ulysses interplanetary field measurements~\citep[][]{wang95}.

Due to the presence of discontinuities in the form of kinks in the field line configuration at the model interface, the coupling of the PFSS and SCS model is not straightforward. As discussed in detail in~\cite{mcgregor08}, we follow the authors' conclusions to reduce the discontinuous behaviour of the magnetic field across the interface of the PFSS and SCS models. Specifically, using a more flexible coupling of the models we set the radius of the source surface to $2.6 \,R_0$ and use the PFSS solution at $2.3 \,R_0$ as an inner boundary condition for the SCS model. This implies that the PFSS model boundary starts at $1 \,R_0$ and extends to $2.6 \,R_0$, and the SCS model boundary starts at $2.3 \,R_0$ and extends to $5 \,R_0$. In this way, we compute the global magnetic field configuration of the solar corona.

From the topology of the solar coronal magnetic field, we compute the solar wind conditions near the Sun that determine the inner boundary condition for solar wind models of the heliosphere. These inner boundary conditions depend crucially on the error in the solar wind speed solution. Empirical formulae for specifying the solar wind speed near the Sun rely on magnetic features computed from the configuration of open magnetic field lines. The areal expansion factor 
\begin{align}
f_{\text{p}} = \left( \frac{R_0}{2.5 \,R_0} \right)^2 \frac{\left|\textbf{B}(R_0,\theta_0,\phi_0)\right|}{\left|\textbf{B}(2.5 \,R_0,\theta_1,\phi_1)\right|},
\end{align}
describes the amount by which a flux tube expands between the photosphere and some reference height in the corona~\citep{wang90}.  Two definitions of the flux tube expansion factor are commonly in use in the scientific literature. One definition is based on the radial component of the magnetic field~\citep[see, for example,][]{riley15}, and one definition is based on the magnitude of the magnetic field~\citep[see, for example,][]{wang90}. We speculate that expansion factor based on the radial component might be more sensitive to the topology of the magnetic flux tube, in particular, at the transition of open and closed field topology, namely coronal hole boundaries. To serve as a point of reference for future investigations, we decided to use the more traditional definition of the flux tube expansion factor, according to~\citet{wang90}.

Specifically, we trace the magnetic field back from the outer boundary of the SCS model to the solar surface, and use the coupled PFSS/SCS model results to compute $f_p$ between 1 and 2.5 solar radii~\citep{wang90,arge00}. In contrast, the great-circle angular distance $d$ refers to the distance between open field footpoints and the nearest coronal hole boundary. It is based on the idea that the solar wind is slow near coronal hole boundaries and fast inside regions of open magnetic field topology~\citep{riley01}.

The most popular techniques for specifying the solar wind speed $v(d,f_\text{p})$ at a reference sphere of $5 \,R_{0}$ (or $30 \,R_{0}$ for the MAS model) are the WS model~\citep{wang90}, the DCHB model~\citep{riley01}, and the WSA model~\citep{arge03}.  The WS relation is based on the inverse relationship between the solar wind speed and the magnetic field expansion factor~\citep{wang90}, namely
\begin{align}
v_{\text{ws}} (f_\text{p}) = a_{1} + \frac{(a_{2} - a_{1})}{{f_\text{p}}^{a_3}}.
\label{eq:eq2}
\end{align}
There is evidence that low magnetic field expansion between the photosphere and some reference height in the corona is correlated with a fast solar wind speed, and vice versa~\citep[e.g.,][]{levine77}. For the coefficients in Equation~\ref{eq:eq2}, we use $a_1=240~\si{km.s^{-1}}$, $a_2=800~\si{km.s^{-1}}$, and $a_3=0.34$, respectively.

The DCHB model correlates the speed at the photosphere with the
distance of an open magnetic field footpoint from the nearest coronal hole 
boundary and maps the calculated solar wind speed solution along the field 
lines to a given reference sphere~\citep{riley01}. The DCHB relation is of 
the form
\begin{align}
v_{\text{dchb}} (d) = b_{1} + \frac{1}{2} \left(b_{2} - b_{1}\right) \left[1 + \tanh \left( \frac{d - b_3}{b_4} \right)\right],
\end{align}
where $b_3$ is a measure for the thickness of the slow flow band, and $b_4$ denotes the width over which the solar wind reaches coronal hole values. For an open field footpoint located at the coronal hole boundary, the solar wind speed is equal to $b_1$. For a footpoint located deep inside a coronal hole, the solar wind speed is equal to $b_2$. This means that the farther away the footpoint is from the coronal hole boundary the faster the expected solar wind speed. In this study, we use $b_1=250~\si{km.s^{-1}}$, $b_2=750~\si{km.s^{-1}}$, $b_3=0.14~\si{rad}$, and $b_4=0.07~\si{rad}$ as the default model parameters.

Finally, the WSA model is a combination of the WS model and the DCHB model that unifies the expansion factor computed from the topology of the magnetic field and the distance from the coronal hole boundary~\citep{arge03}. The WSA relation for specifying solar wind speed near the Sun is given by 
\begin{align}
v_{\text{wsa}} (f_\text{p}, d) = c_1 + \frac{c_2}{\left(1 + f_\text{p}\right)^{c_3}} \left\{ c_4 - c_5 \ \exp \left[ {-\left(\frac{d}{c_6}\right)}^{c_7}\right] \right\}^{c_8},
\label{eq:eq4}
\end{align} 
where $c_i$ are model coefficients. For the coefficients in Equation~\ref{eq:eq4} we use the following settings, $c_1=250~\si{km.s^{-1}}$, $c_2=650~\si{km.s^{-1}}$, $c_3=0.29$, $c_4=1$, $c_5=0.8$, $c_6=3^{\circ}$, $c_7=1.75$ and $c_8 =3$. Finally, we use these empirical techniques as input for models of the heliosphere that propagate the large-scale solar wind solutions from $5 \,R_{0}$ to 1~au. For a more detailed description of the present numerical framework for modeling the evolving ambient solar wind, we would like to refer the reader to~\citet{reiss19} and references cited therein.

\section{The Tunable Heliospheric Upwind Extrapolation (THUX) model} \label{sec:s3}

Several heliospheric models have been proposed to map the solar wind solutions near the Sun to Earth using the coronal model solutions as a boundary condition, each with their own strengths and limitations. The broad spectrum of numerical techniques includes ballistic approximations where each parcel of plasma is assumed to travel with a constant speed through the interplanetary space, to more sophisticated global heliospheric MHD models which attempt to cover all relevant dynamical processes~\citep[e.g.,][]{riley11c,odstrcil03}. In an attempt to optimize the tradeoff between accuracy and processor requirements,~\citet{riley11b} developed the Heliospheric Upwind eXtrapolation (HUX) model by simplifying the fluid momentum equation as much as possible. The authors proposed to neglect the pressure gradient and the gravitation term in the fluid momentum equation to obtain the inviscid form of the one-dimensional Burgers equation defined as
\begin{eqnarray}
- \Omega_\text{rot} \frac{\partial v_r}{\partial \phi} + v_r \frac{\partial v_r}{\partial r} = 0, 
\label{eq:5}
\end{eqnarray}
where $v_r$ is the radial solar wind speed, $\phi$ is the Carrington longitude, and $\Omega_\text{rot}$ is the rotation period of the Sun ($\Omega_\text{rot}=1.53 e^{-4}~\si{deg.s^{-1}}$). Using a forward difference scheme in radial and longitudinal direction, the above equation can be rewritten as
\begin{eqnarray}
- \Omega_\text{rot}  \left( \frac{v_{k+1}^n - v_k^n}{\Delta \phi} \right) + v_k^n \left( \frac{v_{k}^{n+1} - v_k^n}{\Delta r} \right) = 0,
\label{eq:6}
\end{eqnarray}
where the indices $n$ and $k$ denote the radial and longitudinal grid cells, respectively. As outlined in \cite{riley11b}, we can rewrite the above equation to obtain
\begin{eqnarray}
v_k^{n+1} = v_k^n + \frac{\Delta r \, \Omega_\text{rot}}{v_k^n} \left( \frac{v_{k+1}^n - v_k^n}{\Delta \phi} \right).
\end{eqnarray}

When reaching the transition of the coronal/heliospheric model domain, the solar wind plasma has been significantly accelerated towards its final asymptotic speed. However, in the inner heliosphere beyond the coronal model domain, a residual acceleration of the plasma is expected~\citep{schwenn90}. To account for this effect, we follow the approach as discussed in~\cite{riley11b} and simulate the expected residual acceleration by an acceleration term written as
\begin{eqnarray}
v_{\text{acc}}(r) = \alpha \ v_{\text{sw}} \left( 1- \exp\left[-\frac{r}{r_\text{H}}\right]\right),
\end{eqnarray}
where $v_{sw}$ is the computed solar wind bulk speed at the outer boundary of the coronal domain as defined in Equation~\ref{eq:eq2}--\ref{eq:eq4} in Section~\ref{sec:s2}, $\alpha$ is a model coefficient which determines the expected acceleration, and $r_\text{H}$ is a scale length over which the acceleration is expected~\citep{riley11b}. For the model coefficients in this study, we use $\alpha = 0.15$ and $r_\text{H}=50 \, R_0$. Finally, we add the residual acceleration as a function of the heliospheric distance to the model results. 

The established HUX model approach has the advantage that it can match the dynamical evolution explored by global heliospheric MHD codes while having only low computational requirements~\citep{owens17}. This implies that the model is ideally suited for an application in the field of data assimilation to optimally combine the model output with solar wind observations~\citep[see,][]{lang19}. However, previous research comparing observations and ambient solar wind model predictions have found that the complex evolution of wind flows in the interplanetary space results in uncertainties in the predicted arrival times of high-speed streams of more than 24~hours~\citep[see, for instance,][]{reiss16}. For this reason, we propose a modification of the original HUX model which we call Tunable HUX (THUX) model. The THUX model solves the viscous form of the underlying Burgers equation. By adding an additional term on the right-hand side of Equation~\ref{eq:5}, we obtain the viscous form of the Burgers equation given by
\begin{eqnarray}
- \Omega_\text{rot} \frac{\partial v_r}{\partial \phi} + v_r \frac{\partial v_r}{\partial r} = \eta \frac{\partial^2 v_r}{\partial \phi^2},
\label{eq:10}
\end{eqnarray}
where the right-hand side represents the resistance to deformation in propagating the solar wind solutions from the Sun to Earth. Using a central difference scheme for the second derivative on the right-hand side, we can write Equation~\ref{eq:10} on a discretized grid as 
\begin{eqnarray}
- \Omega_\text{rot}  \left( \frac{v_{k+1}^n - v_k^n}{\Delta \phi} \right) + v_k^n \left( \frac{v_{k}^{n+1} - v_k^n}{\Delta r} \right) - \eta \left( \frac{u_{k+1}^n - 2 \, u_k^n + u_{k-1}^n}{\Delta \phi^2} \right) = 0.
\end{eqnarray}

\begin{figure*}
\begin{center}
\includegraphics[width=0.9\columnwidth]{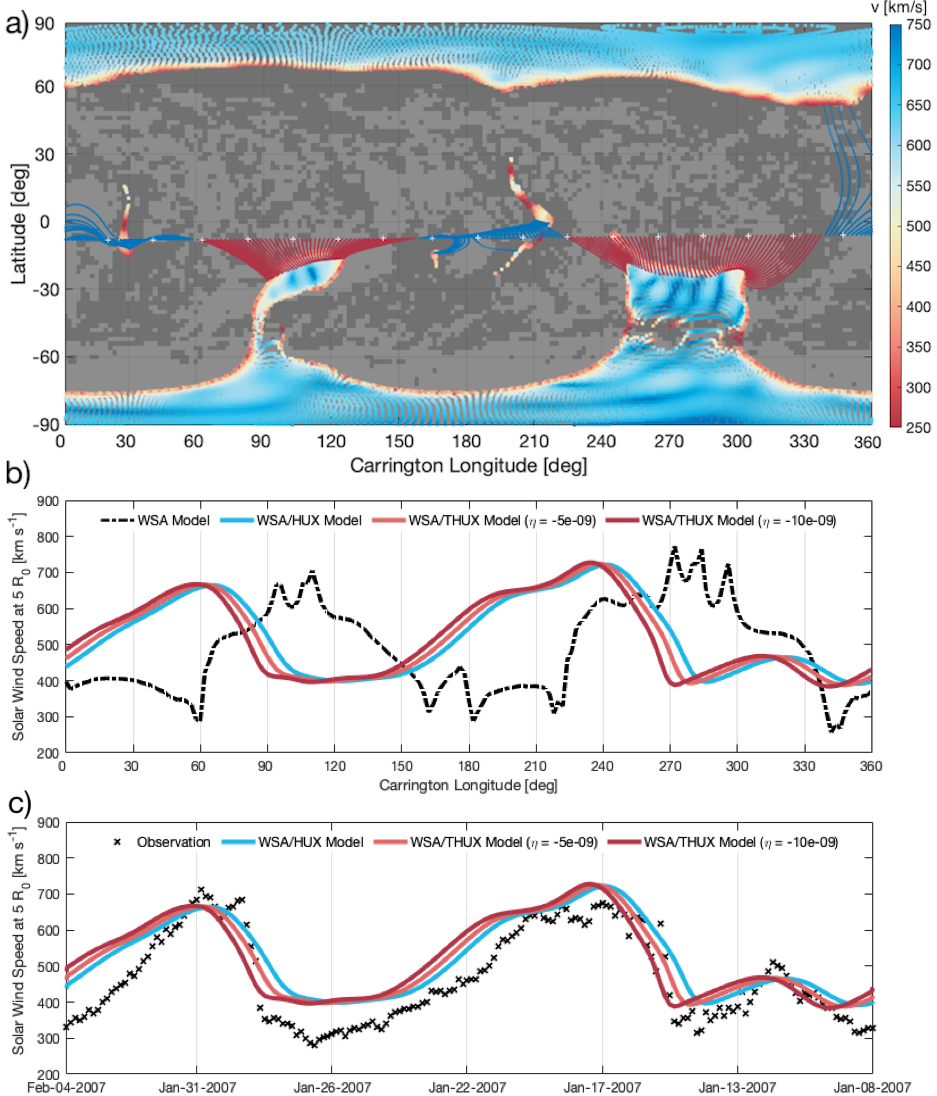}
\end{center}
\caption{Illustration of the coupled WSA/THUX model for predicting the solar wind speed at Earth for the example of CR2052 (i.e., 2007 January 08--2007 February 4). (a) Solar wind speed computed at 5~$R_0$ using the WSA model and mapped along the magnetic field lines to 1~$R_0$. The grey-coloured pixels indicate closed field lines with negative (dark grey) and positive (light grey) magnetic polarity. The field lines show the photospheric footpoints connected to the sub-Earth points at the inner boundary of the heliospheric model at 5~$R_0$. (b) Solar wind speed computed at 5~$R_0$ and propagated to Earth/L1 using the HUX model and the THUX model for two different $\eta$ values; (c) Comparison of the ACE/SWEPAM measurements (black crosses) and the predicted solar wind speed at Earth/L1.
\label{fig:f2}}
\end{figure*}
Transforming the above equation then yields an expression for the solution of the THUX model written as
\begin{eqnarray}
v_k^{n+1} = v_k^n + \frac{\Omega_\text{rot} \, \Delta r}{v_k^n} \left( \frac{v_{k+1}^n - v_k^n}{\Delta \phi} \right) + \frac{\eta \, \Delta r}{v_k^n} \left( \frac{u_{k+1}^n - 2 \, u_k^n + u_{k-1}^n}{\Delta \phi^2} \right),
\end{eqnarray}
where $\eta$ is a model coefficient that controls the resistance to deformation in the solar wind solution. In Figure~\ref{fig:f2}, we illustrate the process of mapping the solar wind solutions near the Sun to the vicinity of Earth using the coupled WSA/HUX and WSA/THUX model. Figure~\ref{fig:f2}(a) shows the solar wind speed solution computed at 5~$R_0$ and mapped along the magnetic field lines to the photosphere, as well as the photospheric footpoints connected to the sub-Earth points at the inner boundary of the heliospheric model at 5~$R_0$. To match the spatial resolution with the resolution of the coronal model, we use $\Delta \phi = 2\degree$ and $\Delta r = 1 \,R_0$, respectively. Figure~\ref{fig:f2}(b) shows the initial solar wind solutions at the sub-Earth points (dashed black line) and the solar wind solutions for different $\eta$ values at $215 \,R_0$ (or 1~au). Figure~\ref{fig:f2}(c) compares the predicted solar wind speed timelines at Earth using the WSA/HUX model and the WSA/THUX model for two different $\eta$ values with in situ measurements of the ambient solar wind at Earth by the Solar Wind Electron Proton and Alpha Monitor~\citep[SWEPAM;][]{mccomas98} onboard the Advanced Composition Explorer~\citep[ACE;][]{stone98} for CR2052. While maintaining the benefits of the HUX model including the low processor requirements~\citep[see, for instance,][]{riley11b,owens17}, the THUX model gives us the possibility to better account for the changing conditions in the ambient solar wind throughout the solar activity cycle. In particular, it is well suited to examine a large number of initial conditions and to deduce error bounds in the context of an adaptive prediction system for specifying solar wind conditions near the Sun.

\section{An adaptive prediction system for specifying solar wind speed} \label{sec:s4}

We present an adaptive prediction system to align the large-scale properties of solar wind speed at the inner boundary of the heliospheric model with the prevailing physical conditions in the vicinity of Earth. The description of the adaptive prediction system is divided into three steps. In Section~\ref{sec:s41}, we present a sensitivity analysis of the WSA model for specifying the solar wind speed near the Sun. We identify the model coefficients that have the lowest impact on the model output and deduce the ranking of the coefficients. By removing these coefficients from the adaptive prediction scheme, we reduce the computational complexity and processor requirements as much as possible. We note that this step is particularly important for the application of the proposed methodology in the context of real-time operational solar wind predictions. In Section~\ref{sec:s42}, we study robust and efficient strategies for creating an ensemble of solar wind solutions; and in Section~\ref{sec:s43}, we select the optimum model coefficients for the adaptive prediction system and deduce confidence boundaries of the solar wind prediction.

\subsection{Sensitivity analysis} \label{sec:s41}
Sensitivity analysis quantitatively assesses how the variation in the output of numerical models can be attributed to variations of the model input~\citep[][]{pianosi16}. The last two decades have seen a growing trend towards sensitivity analysis in the space weather forecasting and modelling community for assessing the robustness of the model results to uncertain inputs and model assumptions. More recently, model developers have shown an increased interest in using sensitivity analysis to study ensembles of solar wind predictions from perturbed initial conditions~\citep[see, for instance,][]{riley13,owens17, linker17}. Although ensemble forecasting has become a very popular technique in the context of solar wind prediction, little attention has been paid to uncertainties in the model settings. In this section, we study the sensitivity of the WSA model on perturbed model settings. For the sake of clarity and consistency, we refer to the areal expansion factor $f_p$ and the distance from the coronal hole boundary $d$ in the described solar wind models as \emph{model parameters} and refer to the model coefficients $(c_1, \dots, c_8)$ as \emph{input factors}.

\begin{figure*}
\begin{center}
\includegraphics[width=0.99\columnwidth]{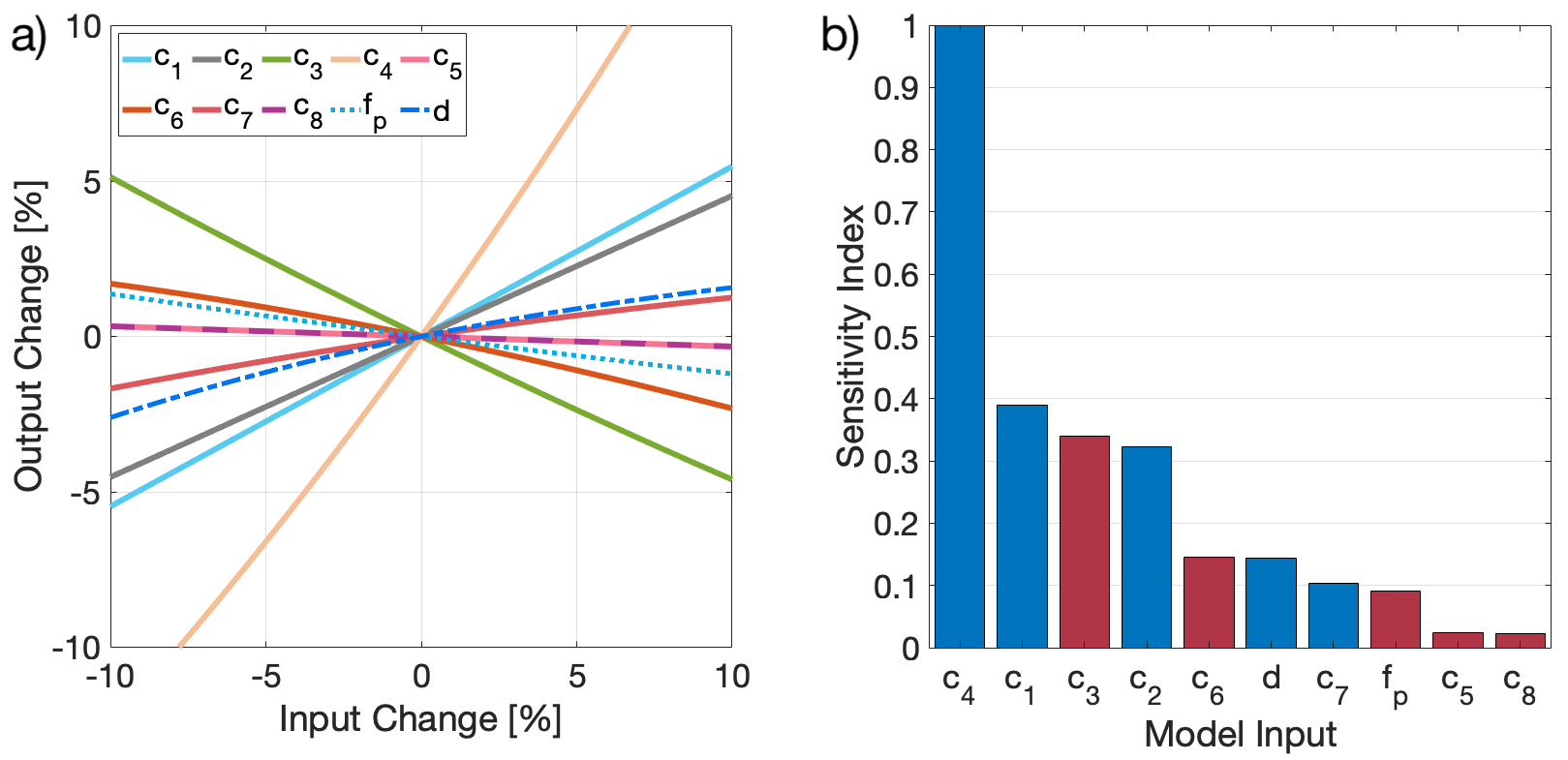}
\end{center}
\caption{Sensitivity analysis of the analytic WSA relation as defined in Equation~\ref{eq:eq4} for specifying solar wind speed near the Sun. (a) Output samples against input samples of different input factors for the WSA model relation with a step size of 0.1~percent; (b) Value of sensitivity index $S_i (\bar x)$ for different input factors. The blue and red coloured bars indicate positive and negative correlations, respectively.\label{fig:f3}}
\end{figure*} 

We apply sensitivity analysis to generate a ranking of the input factors and to identify the input factors that have little impact on the WSA model output. Since the functional relation of the WSA model is available in analytic form in Equation~\ref{eq:eq4}, we can apply the simplest type of sensitivity analysis by varying the input factors from their nominal settings one at a time. The solar wind speed computed from the WSA model is given by $y = v_{\text{wsa}}(\mathbf{c}) = v_{\text{wsa}}(c_1, \dots, c_8)$, where $\mathbf{c}$ is the input vector and $c_i$ are the individual input factors. We compute the output sensitivity to the $i$-th input factor by the partial derivative $\partial y / \partial c_i$ at the nominal value of the input factors $\bar c$. Hence, the sensitivity measure for the $i$--th input factor $c_i$ is defined as
\begin{eqnarray}
S_i (\bar c) = \left. p_i \frac{\partial y}{\partial c_i} \right|_{\bar c},
\end{eqnarray}
where $p_i$ is a scaling factor. A more in-depth analysis of this type is to measure the global sensitivity by calculating output perturbations from multiple points $\mathbf{c}^j$ within the input space. A well known approach in this context is the so called Elementary Effect Test (EET) where the mean of $r$ finite differences (also called Elementary Effects) is taken as a measure of global sensitivity. Setting the scaling factors $c_i$ equal to the input factors, the EET is given by
\begin{eqnarray}
\hat{S}_i = \frac{1}{r} \sum_{j=1}^r \left| S_i(\bar{c})\right| = \frac{1}{r} \sum_{j=1}^r \left| c_i \frac{\partial y}{\partial c_i}\right|.
\end{eqnarray}

Intuitively one would expect that high values of $\hat{S}_i$ indicate that the input factor is more relevant to the global sensitivity of the model than other input factors. Figure~\ref{fig:f3}(a) shows the output samples against samples of different input factors. Figure~\ref{fig:f3}(b) shows the results of the EET for the WSA model for specifying solar wind speed at the inner boundary of the heliospheric model domain. We find that the WSA relation as given in Equation~\ref{eq:eq4} is most sensitive to the input factors $c_4$, $c_1$, $c_3$, and $c_2$ (listed in decreasing order) while the influence of the other input factors is considerably smaller. Therefore, we will rely on the adjustment of the mentioned model input factors to create an ensemble of solar wind solutions at the inner boundary of the heliospheric model. By doing so, we minimize the processor requirements of the adaptive prediction system for the WSA model without neglecting decisive information content.

\subsection{Producing a solar wind speed ensemble}\label{sec:s42}

A key scientific goal in the space weather research community is to develop, implement, and optimize numerical models for specifying the large-scale properties of solar wind conditions at the inner boundary of the heliospheric model domain. The numerical models usually rely on empirical formulae for computing solar wind speed. However, the input factors in the literature can vary dramatically depending on the model implementation and input data products~\citep[see, e.g.,][]{arge03, macneice09b, nikolic14,riley15,pomoell18}. Here we propose a study of an ensemble of possible solar wind solutions around the default model input factors to quantitatively estimate the model uncertainties and to deduce confidence boundaries. Our sampling strategy to create an ensemble of solutions is built on $r$ points in the input space, where the starting point is defined by the default input factors as discussed Section~\ref{sec:s2}. We obtain the subsequent points in the input space by modifying one input factor at a time by a fixed amount of $\delta=2$ percent, where we define the set of incremental percentage adjustments to the input factors used to create the ensemble as $L$. Consequently, the number of ensemble members is the number of elements of $L$ to the power of the total number of input factors. As an example, using the set of incremental percentage adjustments $L=[-4,-2,0,2,4]$ for the input factors $c_1$, $c_2$, $c_3$, and $c_4$, we create an ensemble of solar wind solutions including $5^4=625$ ensemble members.

To propagate each ensemble member from $5 \ R_0$ to $215 \ R_0$ (1~au), we use the THUX model with seven different values for $\eta = [-15 e^{-9}, -10 e^{-9}, -5 e^{-9}, 0, 5 e^{-9}, -10 e^{-9}, 15 e^{-9}]~\si{deg^{2}.s^{-1}}$. We note that modifying $\eta$ by steps of $5 e^{-9}~\si{deg^{2}.s^{-1}}$ changes the arrival time of detected high-speed enhancements~\citep[HSE;][]{reiss16} in the solar wind timelines by approximately 7.5~hours. This means that the adjustments of arrival times are approximately between plus/minus 1~day. Using seven different values for $\eta$ implies that the number of individual solar wind solutions at Earth is $625\times7=4375$. Since the computational requirements increase exponentially with the number of input factors, the screening of input factors in Section~\ref{sec:s41} is an essential part of the proposed adaptive prediction system for the WSA model. The procedure of incremental percentage adjustments is the same for both the WS model and the DCHB model. In both cases, all the input factors are varied to produce an ensemble of solar wind solutions. Furthermore, we note that the value range and step size for $L$ and $\eta$ were determined by balancing the tradeoff between computer requirements and predictive capabilities while considering only physics-based solar wind solutions at the inner boundary of the heliospheric model part. Specifically, we studied the global solar wind solutions of selected Carrington rotations to ensure that the adaptive process does not introduce unrealistically low or high solar wind speed solutions at the polar regions while optimizing the near-Sun solar wind solutions only at the sub-Earth trajectory. This point is particularly important when the adaptive approach is coupled with an MHD code in the future. It is important to stress that we do not claim that the discussed parameter settings are superior to others and should become a community standard. However, we find that the proposed methodology in combination with these settings is both robust and fast, and seems to be a reasonable choice in this first implementation of the adaptive approach for specifying solar wind conditions near the Sun.

\subsection{Selecting the optimum input factors and deducing confidence boundaries}\label{sec:s43}

Validation metrics play a central role in space weather research and forecasting to objectively assess the model agreement with measurements, to constantly diagnose and inform model developers about strength and limitations of the model, and to provide a consistent assessment of the model progress over time. Ideally, ambient solar wind models should be able to accurately simulate both the amplitude and pattern of variability in the solar wind timeline. In reality, the importance of the model skill in terms of a point-to-point error analysis or event-based analysis depends on the application, and it is therefore not possible to define a single metric that can express all relevant aspects. In this section, we will discuss the application of two validation metrics, with one focusing on the amplitude and one on the pattern of variability. We note that the selected validation metric in the adaptive system can easily be replaced by another metric depending on the user needs. 

Each member of the ensemble discussed in Section~\ref{sec:s42} can be associated with a summary scalar variable, for example, a measure of the difference between the measurement $m_i$ and each ensemble member $f_i$. As such, the euclidean distance (ED) can be written as
\begin{eqnarray}
\mbox{ED} = \sqrt{\sum_{i=1}^n \left(m_i - f_i \right)^2}.
\end{eqnarray}
The top $N=1000$ solutions from the previous CR are used to update the model input factors for the next CR. Similar to the approach in~\cite{riley17}, all ensemble members are ranked based on the computed Euclidean distance. In this way, we constantly adapt the model coefficients, compute the ensemble median and deduce confidence boundaries for the prediction. We note that the ensemble median is the preferred average measure as the ensembles of solar wind solutions are often skewed, such that the ensemble mean can yield very biased measures of the ensemble average. 

\begin{figure*}
\begin{center}
\includegraphics[width=0.9\columnwidth]{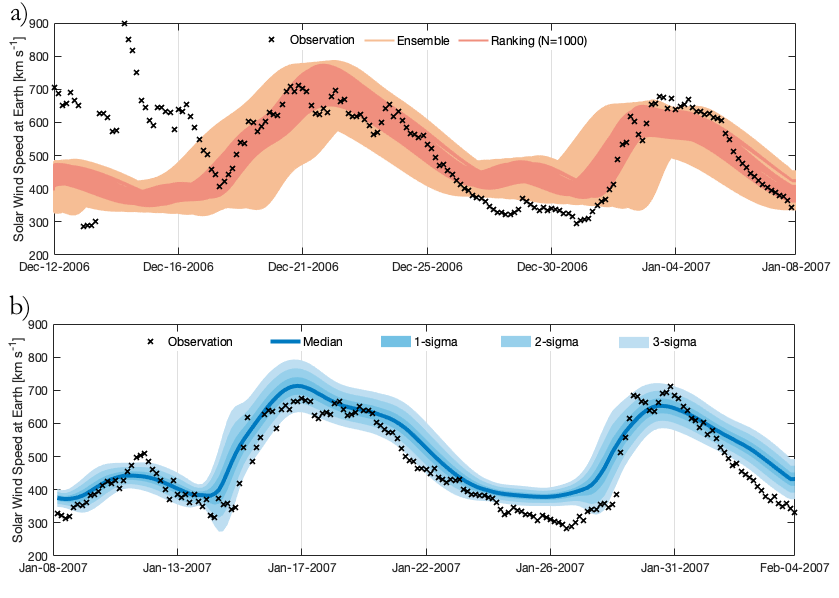}
\end{center}
\caption{Illustration of the adaptive model system on the example of Carrington Rotation 2051--2052 (i.e., 2006 December 12--2007 February 4). (b) Process of training an ensemble of solutions (light-red) to better match the solar wind conditions in the vicinity of Earth (black crosses). The best $N=1000$ solutions (red) are then used to update the model coefficients for the current Carrington Rotation. (a) Comparison of the solar wind speed measurements (black crosses), and the median of the best solutions as deduced from the previous Carrington Rotation (blue line) together with the 1-sigma, 2-sigma, and 3-sigma quantiles.\label{fig:f4}}
\end{figure*}

Additionally, we study a validation metric focusing on the pattern of variability~\citep[see,][]{taylor01}. Given the model variables $f$ and the measurement as a reference $r$, the score $S_0$ is defined as
\begin{align}
S_0 = \frac{4 (1 + C)}{(\hat\sigma_f + 1/\hat\sigma_f)^2(1+C_0)},
\end{align}
where $\hat\sigma_f$ is the ratio of the standard deviations of the model and the observation ($\hat\sigma_f = \sigma_f/\sigma_r$), and $C_0$ is the maximum correlation coefficient that can be expected. For estimating the maximum correlation coefficient we followed the approach discussed in~\citet{taylor01}. It is clear when $\hat\sigma_f$ is equal to 1, that is, when the standard deviation of the model and observation are equal, and $C$ is equal to $C_0$ the score $S_0$ goes to $1$. 

Figure~\ref{fig:f4} illustrates the application of the adaptive prediction system on the example of CR2052. Figure~\ref{fig:f4}(a) shows the training process of CR2052 based on CR2051. The shaded areas indicate the ensemble members as outlined in Section~\ref{sec:s42}, and the red lines indicate the selected solutions based on the computed Euclidean distance. Figure~\ref{fig:f4}(b) shows the median of the ensemble members and the computed error boundaries from the 1000 best model solutions. It is important to note that ensemble averaging of possible future states does not necessarily provide an improvement of deterministic predictions~\citep{jolliffe03, wilks11}. The key advantage of ensemble modelling is the quantitative assessment of prediction uncertainty including error boundaries~\citep[e.g.,][]{owens17}. 

\section{Assessing model quality} \label{sec:s5}

We present a validation analysis of solar wind predictions with in situ measurements of the ambient solar wind at Earth/L1 by ACE/SWEPAM~\citep{stone98,mccomas98} for the time 2006 December 12 -- 2015 September 7. We quantitatively assess the skill of the proposed prediction system using the Operational Solar Wind Evaluation Algorithm~\citep[OSEA;][]{reiss16, reiss19}. OSEA is an open-source Matlab algorithm that runs various predication validation schemes to quantitatively assess the skill of numerical models for predicting the evolving ambient solar wind in the near-Earth environment. 

Traditionally, the relationship between forecast and observation can be studied in terms of continuous variables and binary variables. While the former can take on any real values, the latter is restricted to two possible values such as event/non-event. In the context of solar wind prediction, the solar wind speed time series can be interpreted in terms of both aspects. The forecasting performance can either be evaluated in terms of an average error or its capability in forecasting events of enhanced solar wind speed~\citep[see, for instance,][]{owens05, macneice09a, macneice09b, reiss16}. OSEA is capable of quantifying both aspects, i.e., a continuous variable validation based on simple point-to-point comparison metrics, and an event-based validation analysis assessing the uncertainty of arrival time of high-speed solar wind streams at Earth. 

This section is organized into three parts. First, we present the validation results in terms of established error functions. Secondly, we discuss the forecast performance in terms of binary metrics where each time step in the predicted and observed timeline is labelled as an event/non-event based on the selected threshold value. We summarize the predictive capabilities of the numerical models for a range of event thresholds by the receiver operator characteristic (ROC) curve. Thirdly, we complement our validation analysis by an event-based approach where periods of enhanced solar wind speed, hereinafter referred to as high-speed enhancements (HSE), in predictions and measurements are automatically detected and compared against. 

\subsection{Error functions}
The skill of model predictions of solar wind speed is commonly assessed by error functions such as the root mean square error~\citep[see, for example,][]{wilks11}. To complement error functions of this type, we use the \emph{skill score} (SS) of a prediction expressed as
\begin{eqnarray}
\mbox{SS} = 1 - \frac{{MSE_\text{pred}}}{{MSE}_\text{ref}},
\end{eqnarray}
where $\mbox{MSE}_\text{pred}$ is the mean square error of the predicted timeline, and $\mbox{MSE}_\text{ref}$ is the MSE of a reference baseline model. We use the climatological mean defined as the mean value of the solar wind observation (413.3~\si{km.s^{-1}}) as a reference baseline model~\citep[see, for instance,][]{owens18}. The SS quantifies the improvement over a naive prediction model. Ideally, the prediction skill results in a zero MSE and hence in a SS value of 1. A prediction which equals the skill of the baseline model results in a SS value of 0, and a prediction which is less skilful than the baseline model results in a negative SS value.

\begin{table*}[t]
\caption{The statistical properties of solar wind predictions for CR2047--2167 in terms of arithmetic mean (AM), standard deviation (SD), mean error (ME), mean absolute error (MAE), root mean square error (RMSE), the skill score (SS) relative to the climatological mean, and the Pearson correlation coefficient (PCC). The 27-day persistence model is a baseline reference model against which the prediction models can be compared.}
\begin{center}
\begin{tabular}{lcccccccccc}
Model         & {AM} & SD & ME & MAE & RMSE & SS & PCC\\ 
         & {[}\si{km.s^{-1}}{]} & {[}\si{km.s^{-1}}{]} & {[}\si{km.s^{-1}}{]} & {[}\si{km.s^{-1}}{]} & {[}\si{km.s^{-1}}{]} & \\\hline
WSA				& 469.9	& 82.9		& -56.7	& 91.1		& 111.0		& -0.34& 0.43		\\
Adaptive-WSA(ED)		& 449.5	& 71.9		& -36.2 	& 75.4 		& 93.6 		& 0.05 & 0.50  		\\
Adaptive-WSA($S_0$)		& 480.0	& 85.4		& -66.7 	& 91.7 		& 111.8 		& -0.36 & 0.51   		\\
WS				& 414.3	& 66.4		& 0.1	& 77.0		& 99.4		& -0.27 & 0.29		\\
Adaptive-WS(ED)		& 414.4	& 62.2		& 0.1 	& 69.4		& 90.1     		& -0.06 & 0.41   		\\
Adaptive-WS($S_0$) 	& 419.5		& 65.7			& -4.5	 	& 71.6			& 92.6	     	& -0.25	  & 0.39	   		\\
DCHB			& 432.5 	& 79.7 		& -19.2 	& 81.1 		& 103.3 		& -0.16  & 0.34   		\\
Adaptive-DCHB(ED)		& 428.0	& 75.3		& -14.7	& 71.9		& 92.7		& 0.06 & 0.45   	 	\\
Adaptive-DCHB($S_0$)		& 434.9	& 75.8		& -21.6	& 74.7		& 95.3		& 0.01 & 0.43   	 	\\
Persistence (27-days) 	& 415.1	& 96.5		& -0.5	& 92.7		& 121.6     		& -0.59 & 0.20    \\
%Climatological mean	& 414.5	& 0		& -1.19	& 75.8		& 95.8     		& 0   & 0 \\
Observation  		& 414.5     & 95.7          	& -             & -                   	& -         		& -    & - \\
\end{tabular}
\end{center}
\label{tab:1}
\end{table*}

Table~\ref{tab:1} shows the results obtained from the continuous variable validation of different solar wind models for CR2047 to CR2167 in terms of the arithmetic mean (AM), standard deviation (SD), mean error (ME), mean absolute error (MAE), root mean square error (RMSE), and the skill score relative to the climatological mean (SS). While not strictly an error function, we complement our statistical analysis by the Pearson correlation coefficient (PCC). We study the adaptive approach for ambient solar wind models (WSA, WS, and DCHB model), and run the adaptive system based on the discussed metrics ED and $S_0$ in Section~\ref{sec:s43}. A 27-day persistence model of solar wind speed is a reference baseline model for all the metrics computed in this study against which the predictive capabilities can be compared. Before comparing the observational data with the model solutions at L1, we interpolate the observed time series to be at the same time resolution as our model. Since only the large scale structures in the solar wind solutions are of interest, we select a time resolution of 4~hours for assessing the quality of the model.

We find that the RMSE for the WSA, WS, and DCHB model is 111.0~\si{km.s^{-1}}, 99.4~\si{km.s^{-1}}, and 103.3~\si{km.s^{-1}}, respectively. The fluctuations in the predicted solar wind speeds for the WS model are relatively low ($\text{SD} = 66.4$~\si{km.s^{-1}}). This implies that the WS model benefits greatly in terms of the RMSE for predicting less variability and not for predicting the arrival of enhanced solar wind speeds. The results indicate that the adaptive prediction system using the ED metric improves the capabilities of all three model approaches in terms of the described error functions. For example, the RMSE for the adaptive WSA, WS, and DCHB model using the ED metric is 93.6~\si{km.s^{-1}}, 90.1~\si{km.s^{-1}}, and 92.7~\si{km.s^{-1}} which corresponds to a relative improvement of about 10--15 percent. The same tendency is observable for the correlation coefficient with a relative improvement of about 15--30 percent. Although the PCC also increases by about 15--30 percent, we find that the results for the $S_0$ metric are less promising and provide no improvement for the WSA model in terms of simple point-to-point comparison metrics.

\begin{figure*}
\begin{center}
\includegraphics[width=0.99\columnwidth]{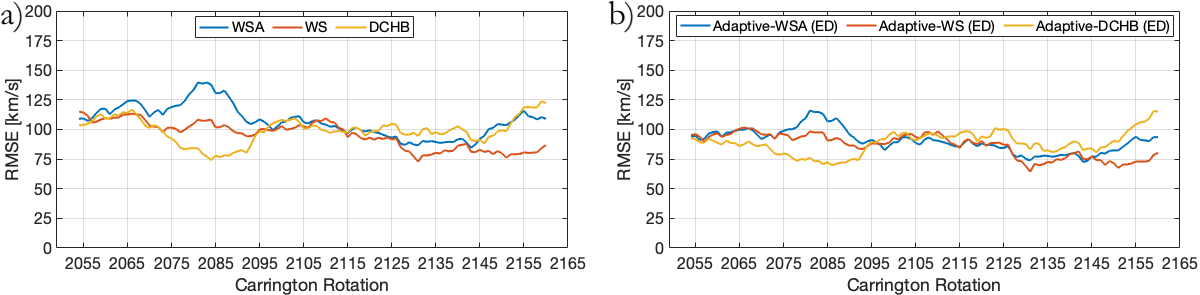}
\end{center}
\caption{Comparison of an 11-Carrington rotation running average of the root mean square error (RMSE) for the WSA (blue), WS (red), and the DCHB model (yellow). (a) Results before the application of the adaptive prediction scheme; (b) Results after the application of the adaptive prediction scheme. \label{fig:movingaverage}}
\end{figure*}

Figure~\ref{fig:movingaverage} shows a comparison of an 11-rotation running average of the RMSE for the WSA, WS, and the DCHB with and without the application of the adaptive prediction scheme. The results indicate that the adaptive prediction scheme generally reduces the RMSE even when the solar activity starts to increase during the year 2010. We also find that the period January 2009--October 2009 (i.e., CR2079--CR2089) is problematic for both the WS model and the WSA model. In comparison, the DCHB model tends to produce better predictions of the solar wind conditions during this interval. We suspect that the presence of unipolar streamers (also known as pseudo-streamers) is artificially raising the baseline of the predicted speed. In context, pseudo-streamers are an important diagnostic for models of the ambient solar wind because they are associated with slow to intermediate speeds despite having very low expansion factors~\citep[e.g.,][]{riley11a}. In general, this characteristic is in contradiction to the basic idea of the WS and the WSA model and the DCHB model thus outperforms these models during the presence of pseudo-streamers.

It should also be noted that all the models produce better results than the reference baseline model (i.e., $\mbox{SS} \ge -0.59$ and RMSE~$\le121.6$). The 27-day persistence model has the same statistics as the measurements and benefits from the quasi-steady and persistent nature of the ambient solar wind flow, especially during the solar minimum phase when polar coronal holes cover a large part of the solar surface. Since our analysis includes times of low and high solar activity, the 27-day persistence model is in reasonable agreement with the observations (e.g., $\text{RMSE} = 121.6$~\si{km.s^{-1}}). Overall, these results indicate that the adaptive model approach using the ED metric improves all the investigated ambient solar wind models, and performs better than the reference baseline model in terms of the computed error functions. 

\subsection{Binary metrics} \label{sec:s52}
While error functions measure the magnitude of the prediction error at every time step, an alternative approach is to consider each time step as an event/non-event. As discussed in~\citet{owens18}, this approach has some advantages over simple point-to-point error measures. First, error functions give equal importance to periods of slow solar wind and periods of fast solar wind, but some users might be interested in the accurate prediction of fast solar wind stream while the detailed evolution of the slow solar wind is of secondary importance. Secondly, outliers in the solar wind time series can have a significant impact on error functions and correlation coefficients. Thus for end-users wanting to react when the solar wind state exceeds a certain threshold, an efficient approach is to consider each time step in the solar wind solution as an event/non-event state. 

In this study, we use an event threshold of $v_{\text{sw}} > 450~\si{km.s^{-1}}$ to define events and non-events in the solar wind timelines. By cross-checking events/non-events in the predicted and observed wind timelines we count the number of hits (\emph{true positives}; TPs), false alarms (\emph{false positives}; FPs), misses (\emph{false negatives}, FNs) and correct rejections (\emph{true negatives}, TNs). From the number of instances, summarized in the so-called contingency table, we compute different skill measures such as the \emph{true positive rate} $\text{TPR} = \text{TP}/(\text{TP} + \text{FN})$, \emph{false positive rate} $\text{FPR} = \text{FP}/(\text{FP} + \text{TN})$, \emph{threat score} $\text{TS} = \text{TP}/(\text{TP} + \text{FP} + \text{FN})$, \emph{bias} $\text{B} = (\text{TP} + \text{FP})/(\text{TP} + \text{FN})$, and \emph{true skill statistics} $\text{TSS} = \text{TPR} - \text{FPR}$. The TSS is defined in the range $[-1,1]$ where a perfect prediction model would have the value 1 (or -1 for a perfect inverse event prediction), and a TSS of 0 indicates no skill. The TSS has the advantage that it uses all elements in the contingency table, and is unbiased by the proportion of predicted and observed events~\citep{hanssen65, bloomfield12}. For further reading on this type of validation measures, we would like to refer the interested reader to~\citet{owens18}. 

\begin{table*}[t]
\caption{Contingency table entries and skill measures of solar wind speed events during CR2047--2167 defined by an event threshold of $v>450$\si{km.s^{-1}}. The table shows the number of Hits ({true positives}; TPs), False Alarms ({false positives}; FPs), Misses ({false negatives}, FNs), Correct Rejections ({true negatives}, TNs), observed ($P$) and forecast events ($P_F$), and observed ($O$) and forecast ($O_F$) non-events. The last three entries in each row show the Threat Score (TS), True Skill Statistics (TSS), and Bias (BS).}
\begin{center}
\begin{tabular}{lccccccccc}
Model         & TP & FP & FN & TN  & TPR & FPR & TS & TSS & BS \\\hline
WSA	& 4265	& 6786	& 1665	& 8227	&	0.72&	0.45&	0.34&	0.27 & 1.86 \\
Adaptive-WSA(ED)	& 4094	& 4961	& 1836	& 10052		& 0.69	& 0.33	& 0.38	& 0.36	& 1.53 \\
Adaptive-WSA($S_0$)	& 4684	& 7159	& 1246	& 7854		& 0.79&	0.48&	0.36&	0.31&	1.99 \\
WS		& 2503	& 3197	& 3427	& 11816 	& 0.42	& 0.21 &	0.27	& 0.21 &	0.96 \\
Adaptive-WS(ED)   	& 2774	& 2627	& 3156	& 12386	& 	0.47& 	0.17& 	0.32& 	0.29&	0.91 \\
Adaptive-WS($S_0$)	& 2896	& 3048	& 3034	& 11965		& 0.49	& 0.20&	0.32&	0.29	&1.00 \\
DCHB		& 2909	& 4343	& 3021	& 10670	& 0.49	& 0.29	& 0.28	& 0.20	& 1.22 \\
Adaptive-DCHB(ED)   	& 3121	& 3268	& 2809	& 11745	& 0.53	& 0.22	& 0.34	& 0.31	& 1.08 \\
Adaptive-DCHB($S_0$)	& 3325	& 3959	& 2605	& 11054		& 0.56	& 0.26	& 0.34	& 0.30	& 1.23 \\
Persistence (27-days)	& 2165	& 3708	& 3728	& 11279		& 0.37	& 0.25	& 0.23	& 0.12	& 1.00 \\
\end{tabular}
\end{center}
\label{tab:2}
\end{table*}

Table~\ref{tab:2} shows the contingency table entries and skill measures of solar wind speed events during CR2047--2167 defined by an event threshold of $v = 450~\si{km.s^{-1}}$. The TSS for the WS, DCHB, and WSA model is 0.21, 0.20, and 0.27 respectively. We find that the TSS for the adaptive prediction system systematically increases. Specifically, the TSS for the newly proposed adaptive WS, DCHB, and WSA model using the ED metric ($S_0$ metric) is 0.29 (0.29), 0.31 (0.30), and 0.36 (0.31), respectively. 

\begin{figure*}
\begin{center}
\includegraphics[width=0.99\columnwidth]{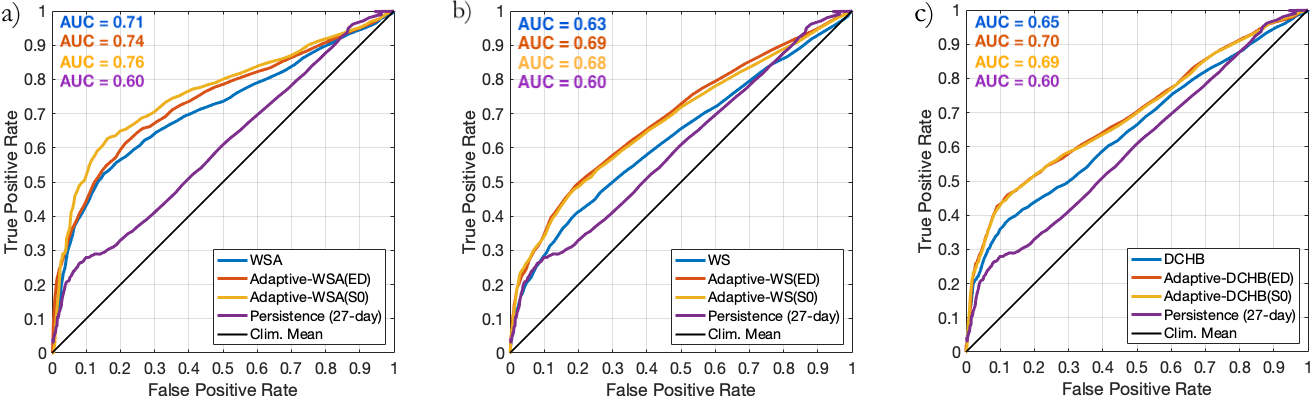}
\end{center}
\caption{Receiver operator characteristic curves plotting the true positive rate (TPR) against the false positive rate (TPR) for a range of event thresholds for different models of the ambient solar wind including the WSA model, WS model, and the DCHB model (a)--(c), together with the 27-day persistence model (violet line) and the climatological mean (black line). The area under the curve (AUC) on the left hand corner is a summary variable of how well the model can predict the ambient solar wind.\label{fig:f5}}
\end{figure*}

The combination of the proportion of correctly predicted events (TPR) and the proportion of falsely predicted events (FPR) in the TSS complement each other and provide deep insight into the capabilities of the model prediction. A way to summarize the predictive capabilities for a range of different event thresholds is the so-called receiver operator characteristic (ROC) curve. The ROC curves illustrate how the number of correctly predicted events (TPR) varies with the number of incorrectly predicted non-events (FPR). Figure~\ref{fig:f5}(a)--(c) shows the resulting ROC curves for the WSA model, WS model, and DCHB model together with the climatological mean. We find that for all the model combinations the results are above the $y=x$ line in Figure~\ref{fig:f5} indicating that $\text{TPR} > \text{FPR}$ for all the ambient solar wind models. A comparison of the different ROC curves shows that the adaptive prediction system improves the results for all the event thresholds. We quantitatively assess this improvement by the computed area under the curve (AUC). The AUC is a summary variable defined between 0 and 1 and an indication of how well the models can predict the ambient solar wind. The analysis shows that the AUC increases for all the adaptive models by approximately 5 percent.

\begin{table*}[t]
\caption{Statistics of the detected high-speed enhancements in terms of event-based metrics including the number of observed (P) and forecast ($\text{P}_\text{F}$) events, the Bias (BS), the number of Hits (TPs), False Alarms (FPs), and Misses (FNs) together with the Probability of Detection (POD), False Negative Rate (FNR), Positive Predictive Value (PPV), False Alarm Ratio (FAR), and Threat Score (TS).}
\begin{center}
\begin{tabular}{lccccccccccccccc}
Model    & P & $P_F$ & BS& TP & FP & FN & POD & FNR & PPV & FAR & TS \\ 
WSA	& 301	& 286 	& 0.92 & 161		& 125	 & 140   & 0.53	& 0.47		& 0.56	& 	0.44 &	 0.38 \\
Adaptive-WSA(ED)		& 301	& 213	& 0.68& 143	& 70	 & 158	& 0.47	& 0.53		& 0.67	& 	0.33 &	 0.39 \\
Adaptive-WSA($S_0$)		& 301	& 246		& 0.78& 159	& 87	 & 142	& 0.53	& 0.47		& 0.65	& 	0.35 &	 0.41 \\
WS				& 301	& 193		& 0.64 & 122	& 71	 & 179	& 0.41		& 0.59		& 0.63	& 	0.37 &	 0.33 \\
Adaptive-WS(ED)			& 301	& 167		& 0.55 & 121	& 46	 & 180	& 0.40		& 0.60		& 0.72	& 	0.28 &	 0.35 \\
Adaptive-WS($S_0$)   		& 301	& 170		& 0.56 & 119	& 51	 & 182	& 0.40		& 0.60		& 0.70& 	0.30 &	 0.34 \\
DCHB			& 301	& 279		& 0.92 & 148	& 131	 & 153	& 0.49		& 0.51		& 0.53	& 	0.47 &	 0.34 \\
Adaptive-DCHB(ED)		& 301	& 206		& 0.68 & 136	& 70	 & 165	& 0.45		& 0.55		& 0.66	& 	0.34 &	 0.37 \\
Adaptive-DCHB($S_0$)		& 301	& 235		& 0.78 & 146	& 89	 & 155	& 0.49		& 0.51		& 0.62	& 	0.38 &	 0.37 \\
Persistence (27-days)	& 313	& 314		& 1.0 & 142	& 172	 & 171	& 0.45		& 0.55		& 0.45	& 	0.55 &	 0.29\\
\end{tabular}
\end{center}
\label{tab:3}
\end{table*}

\begin{figure*}
\begin{center}
\includegraphics[width=0.99\columnwidth]{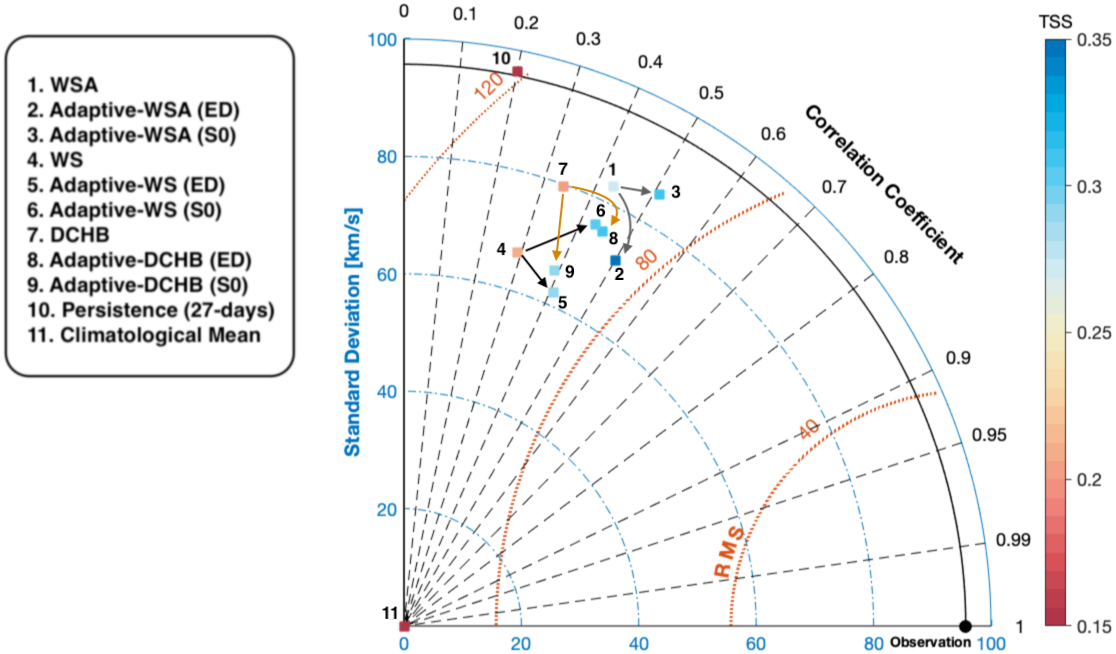}
\end{center}
\caption{Taylor Diagram for displaying a summary of different validation metrics, including the standard deviation, root mean square error (RMSE) and Pearson correlation coefficient (PCC). The diagram compares the predictive capabilities of the WSA model, WS model, DCHB model, Adaptive-WSA model, Adaptive-WS model, Adaptive-DCHB model, persistence model, and the climatological mean with the measured solar wind speed for CR2047--CR2167. The colourmap complements the computed error functions by displaying the TSS calculated from an event/non-event based validation analysis as outlined in Section~\ref{sec:s52}.
\label{fig:f6}}
\end{figure*}

In parallel to that analysis, we use the Taylor diagram to summarize different validation metrics in a single diagram~\citep{taylor01}. In recent years, the Taylor diagram has become a popular means to present multiple aspects of model validation in a single diagram~\citep[e.g.,][]{riley13, owens18}. The key of constructing such a diagram is to recognize the geometric relationship between the correlation coefficient, the RMSE, and the amplitude of variations in the predicted and reference time series~\citep[see,][]{taylor01}. In this way, we compare different model results and trace the impact of the proposed modifications. As shown in Figure~\ref{fig:f6}, the azimuthal position indicates the PCC, the radial distance from the circle at the x-axis is proportional to the RMSE, and the distance from the origin is proportional to the amplitude of variations (standard deviation). Thus, model predictions in good agreement with the observations will be located very close to the circle on the x-axis indicated by similar standard deviation, high correlation and low RMSE. The distance from the circle on the x-axis refers to the overall model performance in terms of the underlying validation metrics. To complement the results displayed in the Taylor diagram, we use a colourmap indicating the TSS value from the discussed event/non-event analysis (see, Table~\ref{tab:2}). Figure~\ref{fig:f6} shows that all the solar wind solutions using the adaptive prediction system are located closer to the observation. The arrows in Figure~\ref{fig:f6} highlight the improvement of the WSA, WS, and DCHB model by the application of the adaptive model approach. 

\subsection{Event-based metrics}
A core difficulty facing any validation analysis that uses error functions as quantitative validation measures is an inadequate knowledge of the uncertainties due to timing errors~\citep[see,][]{owens18}. The interpretation of simple point-to-point error measures can be misleading as discussed in the scientific literature~\citep[see, for example,][]{owens05, macneice09a, macneice09b}. A validation analysis is very challenging, for example, when the evolution of significant leaps in the timelines are generally well predicted, but the arrival times differ slightly in prediction and observation. To account for the uncertainty in the arrival times, we use an event-based validation approach as discussed in greater depth in~\cite{reiss16}. More specifically, the validation analysis applied in this section consists of three steps. First, we define and detect events of enhanced solar wind speed, also called high-speed enhancements (HSEs) in forecast and observation data. Secondly, we associate the HSEs detected in the solar wind measurements with HSEs detected in the prediction and label each event pair as a hit, false alarm, or miss. Thirdly, we compute different validation summary variables to present and compare the predictive abilities of the investigated models.

Table~\ref{tab:3} shows the number of observed ($P$) and forecast ($P_F$) events, the Bias (BS), the number of Hits (TPs), False Alarms (FPs), and Misses (FNs) together with the Probability of Detection (POD), False Negative Rate (FNR), Positive Predictive Value (PPV), False Alarm Ratio (FAR), and Threat Score (TS). With a minimum event speed limit of 450~\si{km.s^{-1}}, the POD for the WSA is 0.53, the FNR is 0.47, the PPV is 0.56, and the FAR is 0.44. This means that about 53 percent of the observed HSEs are correctly predicted by the WSA model, and 47 percent of all predicted HSEs are observed. In contrast, the POD for the Adaptive-WSA model is 0.53, the FNR is 0.47, the PPV is 0.65, and the FAR is 0.35. This indicates that the adaptive prediction system reduces the number of false alarms by about 30 percent. A similar improvement is observable for the other models, with the false alarms for the WS and DCHB being reduced by 25 and 30 percent, respectively. In comparison, the TS for the WSA model is 0.38 whereas the TS for the Adaptive-WSA(ED) is 0.39. Although times of recorded ICMEs according to the list\footnote{\url{http://www.srl.caltech.edu/ACE/ASC/DATA/level3/icmetable2.htm}} of Richardson and Cane~\citep{richardson10} are excluded from our analysis, all ambient solar wind models have in common that they systematically underestimate the number of observed HSEs ($\si{BS} \le 0.92$). To support a consistent and transparent assessment of space weather modeling products, we have uploaded all the discussed validation measures and the updated validation functions to the OSEA online repository\footnote{\url{https://bitbucket.org/reissmar/solar-wind-forecast-verification}}.

\section{Discussion} \label{sec:s6}
In this study, we presented an adaptive prediction system that fuses information from in situ observations of the evolving ambient solar wind flow into numerical models to better match the global solar wind solutions near the Sun with prevailing physical conditions in Earth's space weather environment. A novel element of our procedure is the replacement of static empirical formulae by a more flexible approach for specifying solar wind conditions near the Sun. We coupled the continuously updated solar wind conditions with computationally efficient heliospheric propagation tools such as the HUX model and the newly developed THUX model to deduce the optimum model coefficient settings for the following CR. By doing so, we studied an ensemble of possible solar wind solutions, quantitatively estimated the model uncertainties and deduced confidence boundaries. Finally, we applied a comprehensive validation analysis based on simple point-to-point error measures and event-based measures, and compare our results to reference benchmark models. Our study leads us to two primary conclusions: 
\begin{itemize}
\item[i)] solving the viscous form of the Burgers equation in the newly proposed THUX model provides a robust and efficient method to study, adapt, and optimize the input variability space of established empirical relationships between the magnetic field topology and the near-Sun solar wind conditions including the WS, DCHB and WSA model; and 
\item[ii)] the results show that the application of the proposed adaptive prediction system improves the abilities of models of the ambient solar wind for real-time operational purposes according to the community validation measures applied.
\end{itemize}
\noindent We find that the proposed prediction scheme improves all the investigated coronal/heliospheric model combinations and that both of our conclusions have an essential impact on enhancing the predictive abilities of the model approaches investigated. Although the basic idea of the THUX model was to provide us with an efficient means to study a variety of solar wind model solutions, we find that the application of the THUX without the adaptive process in the empirical speed formulae also improves the results. As an example, we find that in the case of the WSA model this approach reduces the RMSE to 107.8 km/s and increases the CC to 0.45. We can therefore deduce that the application of the THUX alone has a positive impact on the results, even without the application of the full adaptive prediction scheme. Furthermore, it should be noted that our results for the WSA model with the default model settings, are in reasonable agreement with other studies~\citep[e.g.,][]{owens08,reiss16}. This implies that the described methodology using the default model parameter settings is starting from a similar level of forecasting skill and that application of the adaptive prediction system might be beneficial for other solar wind frameworks too. Another important finding is that the adaptive scheme also improves the results during periods of increased solar activity even when the magnetic field configuration in the solar corona is highly structured, and the dynamical evolution of the evolving ambient solar wind is much more complicated.

It is crucial to bear in mind the possible uncertainties that affect the results and the conclusions of our study. An essential component in this context is the uncertainty of observed magnetic maps, which influence the quality of solar wind solutions with empirical and more physics-based MHD models. Especially the process for constructing synoptic maps and the correction of the poorly observed polar regions likely has a considerable influence on the resulting model solutions~\citep{riley12a}. In view of this, one could question the ability of coronal models for reconstructing the global topology of open magnetic field lines. To seamlessly simulate the dynamics of the evolving ambient solar wind, we rely on the coupling of a magnetic model of the corona with the HUX and THUX model to map the solutions near the Sun to Earth. The coronal part of our numerical framework uses the PFSS model to reconstruct the global magnetic field configuration. Ideally, one would prefer more physics-based models to simulate the complex dynamics of the evolving ambient solar wind, especially at solar wind stream interaction regions, which are not included in the present approach. Nevertheless, recent research suggests that the predictive abilities of semi-empirical and full physics-based coupled corona/heliosphere models are very similar. As an example, \cite{owens08} studied the performance of different numerical frameworks (WSA, WSA/Enlil, and MAS/Enlil) and showed that the coupled semi-empirical approach gives the best results in point-to-point measures. To optimize the tradeoff between model accuracy and processor requirements of a full MHD code, we would recommend usage of the described methods, in particular, for the purpose of real-time solar wind predictions. 

In the context of real-time operational solar wind prediction, it is essential to highlight that we used Carrington rotation magnetic maps to present the application of this prototype of an adaptive scheme. This means that the sub-Earth observations of the photospheric magnetic field are between 0 and 27 days old. Hence, the present framework for prediction does not correspond to real-time operational space weather frameworks for prediction of the solar wind state in the interplanetary medium and at Earth's space weather environment. Although an analysis of real-time applications is beyond the scope of this study, we speculate that usage of the most recent magnetic maps in combination with the adaptive prediction scheme with a daily update could also enhance the accuracy of real-time wind predictions. 

We want to emphasize that we have conducted some additional experiments and investigated various model settings to optimize the results of the adaptive scheme which are not mentioned in this study. As an example, we studied the impact of the adaptive process on the large-scale solution of the solar wind models near the Sun to exclude non-physical solutions from the beginning. In this way, we were able to reject model settings that produced unrealistically low or high values of the solar wind speed at the polar regions. Moreover, we implicitly dealt with the problem of overfitting by allowing only for variations in the model coefficients up to a maximum of 4 percent. In addition, we studied different time windows for the adaptive process. Specifically, we increased the size of the training window up to 5 Carrington rotations but saw no significant improvement to the model results during this process. Along these lines, it is important to note that we do not claim that the present parameter settings should become a community standard. Since the framework implementation could rely on different full-disk magnetograms and could be very technically different, it is not guaranteed that the settings discussed here are suitable for all operational solar wind prediction pipelines. Instead, this study presents the first prototype of such an implementation and we recommend that the parameter settings should be investigated based on the underlying numerical framework. 

In context, we would like to note that the adaptive approach discussed in this work is not guaranteed to make the global three-dimensional solar wind more accurate. The three-dimensional boundary conditions at the inner part of global heliospheric codes provide the solar wind speed on a full sphere, with the latitudinal structure of the high-speed streams are important in determining whether a stream will reach Earth, or for the dynamics of coronal mass ejections that propagate close to the streams. In this study, the adapted model coefficients are only trained to fit the near-Earth observations and have not been tested to derive boundary conditions for three-dimensional heliospheric models. In the future, however, we plan to work on different approaches for coupling the proposed adaptive boundary conditions with a global heliospheric code.

In closing, we note that there are further benefits to this approach, of which one is the improvement of deterministic forecasting by including a quantification of the associated uncertainty. Future work will look towards applying more sophisticated statistical methods to improve the adaptive process where possible, and will also look to extending the model through the inclusion of further physical properties beyond the solar wind speed. Furthermore, this approach shows promise in the application to coronal model parameters such as the source surface distance. Recent studies on the source surface distance suggest that the choice of the distance influences the results largely, and this varies with the solar activity~\citep[e.g.,][]{lee11,arden14,nikolic19}. A problem here is that of choosing the right model coefficients in Equation~\ref{eq:eq2}--\ref{eq:eq4} for the solar wind models for the relevant source surface height. With the adaptive process method described here, we suggest that it might be possible to derive continuously updated boundary conditions near the Sun for the corresponding source surface height. Improvements to the boundary conditions of heliospheric models represent not only an improvement for space weather prediction but also for space weather research in general. Therefore we conclude that our study has important implications for future work in applied space weather research and prediction.

\section*{Acknowledgments}
The work utilizes data obtained by the Global Oscillation Network Group (GONG) Program, managed by the National Solar Observatory, which is operated by AURA, Inc.~under a cooperative agreement with the National Science Foundation. The data were acquired by instruments operated by the Big Bear Solar Observatory, High Altitude Observatory, Learmonth Solar Observatory, Udaipur Solar Observatory, Instituto de Astrof\'{i}sica de Canarias, and Cerro Tololo Interamerican Observatory. The authors thank Yi-Ming Wang and Leila M.~Mays for helpful conversations about this work, and acknowledges NASA's Community Coordinated Modeling Center (CCMC) for financial travel support. K.M. acknowledges support by the NASA HGI program (\#80HQTR19T0028). M.A.R., C.M., R.L.B., A.J.W., T.A. and J.H~acknowledge the Austrian Science Fund (FWF): J4160-N27, P31659-N27, P31521-N27, and P31265-N27.

\bibliographystyle{unsrt}  


\end{document}